\documentclass[usegraphicx,usenatbib]{mn2e}
\usepackage{txfonts}
\usepackage[Symbol]{upgreek}
\usepackage[T1]{fontenc}
\usepackage[utf8]{inputenc}

\title[Radio galaxy J0028$+$0035]{Multifrequency study of a double--double 
radio galaxy J0028$+$0035}
\author[A. Marecki, M. Jamrozy, J. Machalski, \& U. Pajdosz-Śmierciak]
{A. Marecki$^{1}$\thanks{E-mail: \texttt{amr@astro.uni.torun.pl}}, M. Jamrozy$^{2}$, J. Machalski$^{2}$, U. Pajdosz-Śmierciak$^{2}$\\
$^{1}$Institute of Astronomy, Nicolaus Copernicus University, Faculty of Physics, Astronomy and Informatics, ul. Grudziądzka 5,
PL-87-100 Toruń, Poland\\
$^{2}$Astronomical Observatory, Jagiellonian University, ul. Orla 171, PL-30-244 Kraków, Poland}

\begin{document}

\date{Accepted 2020 November 16. Received 2020 November 16; in original form 2020 July 31}

\pagerange{\pageref{firstpage}--\pageref{lastpage}} \pubyear{0000}

\maketitle

\label{firstpage}

\begin{abstract}

We report the discovery of a double--double radio source (DDRS) J0028$+$0035.
We observed it with LOFAR, GMRT, and the VLA. By combining our observational 
data with those from the literature, we gathered an appreciable set of radio 
flux density measurements covering the range from 74\,MHz to 14\,GHz. This 
enabled us to carry out an extensive review of physical properties of the 
source and its dynamical evolution analysis. In particular, we found that, 
while the age of the 
large-scale outer lobes is about 245\,Myr, the renewal of the jet activity, 
which is directly responsible for the double--double structure, took place 
only about 3.6\,Myr ago after about 11\,Myr long period of quiescence.
Another important property typical for DDRSs and also present here is 
that the injection spectral indices for the inner and the outer pair of 
lobes are similar. The jet powers in J0028$+$0035 are similar too.
Both these circumstances support our inference that it is, in fact, a DDRS which was 
not recognized as such so far because of the presence of a coincident 
compact object close to the inner double so that the centre of J0028$+$0035 
is apparently a triple.

\end{abstract}


\section{Introduction}
\label{intro}

Intermittent nature of the activity is a well-known property of active galactic 
nuclei (AGN). The most compelling observable signature of cessation and 
subsequent restart of activity in a radio-loud AGN takes the form of
a double--double radio source (DDRS) that consists of two co-linear pairs
of radio lobes \citep{Schoenmakers2000}. The outer lobes develop during the 
earlier active period. Once the active nucleus becomes quiescent, the supply 
of relativistic particles via jets to the lobes is cut off, and they
gradually fade out. Quiescence can last for up to $10^8$\,yr
\citep{Komissarov1994}. Nevertheless, having not been fuelled even for such
a long time, the lobes may still be visible; they usually appear as diffuse
relics. Meanwhile, the renewal of activity of the nucleus may occur, which
leads to the creation of a new pair of lobes -- the inner one. The 
possibility that they emerge \textit{before} the outer lobes 
disappear is the reason why the overall radio structure is double--double.

Nearly a hundred of DDRSs are known to date \citep[see 
e.g.][]{Saikia2009, Nandi2012, Kuzmicz2017, Mahatma2019}. Some of them 
have unusual properties. There are three objects with clear evidence of 
possessing even three pairs of lobes, so they are labelled triple-double 
radio sources. These are: J0929$+$4146 
\citep{Schoenmakers2000,Brocksopp2007}, J1409$-$0302 
\citep[Speca;][]{Hota2011}, and J1216$+$0709 \citep{Singh2016}. Two DDRSs,
J0935$+$0204 \citep[4C\,02.27;][]{Jamrozy2009} and J0741$+$3112 
\citep{Siemiginowska2003}, are hosted by quasars, whereas J2345$-$0449 
\citep{Bagchi2014} and again J1409$-$0302 -- both of giant sizes -- are 
hosted by spiral galaxies. Another exceptional case of restarting activity 
is the giant radio galaxy J2333$-$2343 \citep{Hernandez2017} where the 
direction of the jet has changed in the new episode of nuclear activity; it 
points almost towards us now which makes this source a radio galaxy with a 
blazar host. J1647$+$4950 \citep[SBS\,B1646$+$499;][]{Pajdosz2018} with 
large-scale diffuse lobes is also hosted by a blazar.

\begin{figure}
\includegraphics[width=1.0\linewidth]{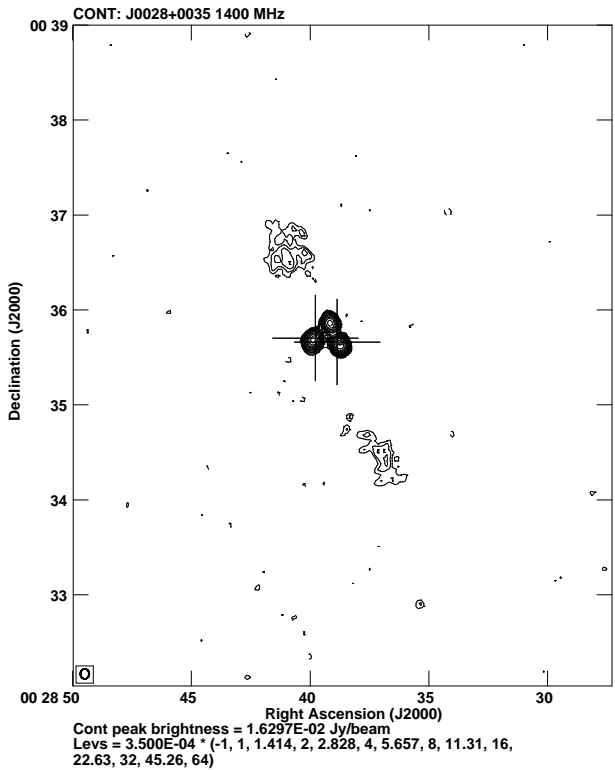}
\caption{FIRST image of J0028$+$0035. The optical position of the
BL\,Lac object 5BZU\,J0028$+$0035 is marked with the left-hand cross. The position of
SDSS\,J002838.86$+$003539.7 galaxy is marked with the right-hand cross.}
\label{fig:FIRST}
\end{figure}

Surprisingly, a relatively small number of DDRSs were thoroughly studied. 
These are: 1548$-$3216 \citep{Machalski2010}, J1453$+$3308 
\citep{Konar2006}, J0041$+$3224 and J1835$+$6204 \citep{Konar2012, 
Konar2013}, J1352$+$3126 \citep{Machalski2016}, and J1706$+$4340 for which
\cite{Marecki2016} determined a range of astrophysical parameters such as 
dynamical and synchrotron ages, and the ambient medium characteristics. In 
particular, they found that the large-scale outer lobes of J1706$+$4340 are 
up to 300\,Myr old whereas the renewal of the jet activity took place only 
about 12\,Myr ago after about 27-Myr long period of quiescence. This result 
is yet another but very compelling proof that the explanation of the nature 
of DDRSs proposed by \cite{Schoenmakers2000} is correct. The correctness of 
their model can also be demonstrated when studying a sample of DDRSs 
selected from a large survey like \textit{Faint Images of the Radio Sky at 
Twenty} centimetres \citep[FIRST;][]{Becker1995}. This has been carried out 
in three steps. The initial selection was made by \cite{Proctor2011} who 
found 242 DDRS candidates. Secondly, 23 of them were verified as true DDRSs by 
\cite{Nandi2012}. Finally, the sample was surveyed with the Giant Metrewave
Radio Telescope \citep[GMRT,][]{Swarup1991} by \cite{Nandi2019}. 
Using both 607-MHz and the FIRST images, they estimated spectral indices of 
the inner and the outer lobes of each source. For the vast majority of them, 
the spectra of their outer lobes were significantly steeper in line
with \cite{Schoenmakers2000}.

When searching FIRST for the large-scale structures centred on the objects 
from Roma-BZCAT, a multifrequency catalogue of blazars \citep{Massaro2009}, 
we discovered that the BL\,Lac object 5BZU\,J0028$+$0035, also known as
SDSS\,J002839.77$+$003542.2, not only was straddled by large-scale relic lobes but also 
accompanied by two other compact components located close enough to it (in 
terms of the angular distance) to be straddled by those relics as well. We 
suggest that the only way to interpret the triple structure at the centre of 
J0028$+$0035 -- see Fig.\,\ref{fig:FIRST} -- is to assume that 5BZU\,J0028$+$0035
is a coincidence. It follows that since the relic lobes and the inner two 
components at RA\,$=\!00^{\rm h} 28^{\rm m} 38\fs76$, Dec.\,$=\!+00\degr 
35\arcmin 38\farcs1$ and RA\,$=\!00^{\rm h} 28^{\rm m} 39\fs16$, 
Dec.\,$=\!+00\degr 35\arcmin 51\farcs6$ are co-linear, 
J0028$+$0035 could be a DDRS. In other words, it appears 
that this is because of the presence of 5BZU J0028$+$0035 that 
J0028$+$0035 was not recognized as a DDRS so far. To support the 
classification of J0028$+$0035 as a DDRS, its optical identification and 
higher resolution observations of its putative inner lobes are needed. 
Obviously, the host galaxy must lie between them. 
SDSS\,J002838.86$+$003539.7 is the only candidate here; it is marked with 
the right-hand cross in Fig.\,\ref{fig:FIRST}. As for the radio morphology 
of the alleged inner lobes, revealing it is desirable because they are 
unresolved in the FIRST image (Fig.\,\ref{fig:FIRST}). Therefore, based on 
that image alone, it is not possible either to rule out the 
possibility that the inner triple is a result of e.g. gravitational 
lensing or to confirm whether the morphologies of the two western 
components are typical for the lobes. Owing to the 
observations we have carried out, we are able to address 
this issue. Furthermore, we analyse all available radio data on J0028$+$0035
in order to study its dynamics and energetics.

The paper is organized as follows. Multifrequency radio data are presented 
in Section\,\ref{radio_data}. The dynamical analysis of each pair of radio 
lobes of J0028$+$0035 is elaborated upon in Section\,\ref{dyn_evo}. The 
results are discussed in Section\,\ref{discussion} and summarized in 
Section\,\ref{summary}. For consistency with the analyses conducted in our 
earlier papers, the cosmological parameters by
\citet[][$H_0\!=\!71\,{\rm km\,s}^{-1}{\rm Mpc}^{-1}, \Upomega_{\rm 
M}\!=\!0.27$, $\Upomega_{\Uplambda}\!=\!0.73$]{Spergel2003} are used 
throughout this article. Positions are given in the J2000.0 coordinate 
system.

\begin{table*}
\caption{Radio data of J0028$+$0035 analysed in this study}
\begin{tabular}{r c l r r r l c}
\hline
Frequency  & Telescope/         & \multicolumn{1}{c}{Date of} & \multicolumn{2}{c}{Beam size}             &  Beam PA  & \multicolumn{1}{c}{rms} & Reference \\
\multicolumn{1}{c}{(MHz)} & Survey             & \multicolumn{1}{c}{observation} & \multicolumn{1}{c}{(arcsec)} & \multicolumn{1}{c}{(arcsec)} & ($^\circ$) &(mJy beam$^{-1}$)&      \\
           &                    &               &                     &                     &            &           &      \\ 
\multicolumn{1}{c}{(1)}    &    (2)             & \multicolumn{1}{c}{(3)} & \multicolumn{2}{c}{(4)}         & (5)        &  \multicolumn{1}{c}{(6)}    &  (7) \\ 
\hline
           &            	&            	&                     &                     &            &           &      \\
  73.8     &VLA-B/VLSSr 	& 2003 Sep. 20	&   75                & 75                  & 0          & 180.7     &  1   \\ 
 76--227   &MWA/GLEAM           & 2013 Nov. 15  &   324.2 -- 118.0    & 295.7 -- 106.6      &            & 95 -- 20  &  2   \\  
 143.7     &LOFAR HBA           & 2017 Aug. 18  &   14.54             & 6.25                & 76.1       &   1.0     &  p   \\ 
 147.6     &LOFAR HBA           & 2017 Aug. 18  &   12                & 12                  & 0          &   1.7     &  p   \\
 322.7     &GMRT        	& 2017 Sep. 14 	&   9.43              & 7.79                &74.73       &   0.09978 &  p   \\
 607.7     &GMRT        	& 2017 Sep. 10 	&   5.34              & 4.14                &61.39       &   0.1135  &  p   \\
 607.7$^{a}$&GMRT        	& 2017 Sep. 10	&   12.49             & 7.95                & $-$36.36   &   0.2375  &  p   \\     
1400.0     &VLA-D/NVSS  	& 1993 Nov. 15	&   45                & 45                  & 0          &   0.45    &  3   \\ 
1400.0     &VLA-B/FIRST 	& 1995 Sep. 23  &   6.40              & 5.4                 & 0          &   0.11    &  4   \\ 
1437.4     &GMRT        	& 2017 Sep.  5  &   4.37              & 3.60                & 82.09      &   0.0788  &  p   \\
1519.4     &VLA-A               & 2018 Apr. 18  &   2.47              & 1.01                &$-$50.95    &   0.0895  &  p   \\
3000.0     &VLA-B/VLASS         & 2017 Sep. 29  &   2.76              & 2.16                &$-$7.52     &   0.1033  &  5   \\
4850.0     &Green Bank 91m/87GB	& 1986-1987  	&   222               & 194                 & 0          &   5       &  6   \\ 
5469.1     &VLA-C               & 2018 Nov. 21  &   7.78              & 3.02                &$-$51.77    &   0.0130  &  p   \\
5494.0     &VLA-A               & 2018 Apr. 2   &   0.75              & 0.30                &$-$53.44    &   0.0138  &  p   \\
9016.8     &VLA-D               & 2018 Sep. 12  &   15.87             & 7.21                & 49.36      &   0.0160  &  p   \\ 
14014.9    &VLA-D               & 2018 Sep. 12  &   9.68              & 4.38                & 49.12      &   0.016   &  p   \\

\hline
\end{tabular}
\begin{flushleft}
{\em References} --
(1)~\cite{Lane2014};
(2)~\cite{Hurley-Walker2017}; 
(3)~\cite{Condon1998}; 
(4)~\cite{Becker1995}; 
(5)~\cite{Lacy2020}; 
(6)~\cite{Gregory1991};
(p)~this paper.
$^a$Data in this line has been extracted from the tapered map.
\end{flushleft}
\label{table:radio_data}
\end{table*}

\section{Radio data on J0028$+$0035}
\label{radio_data}

Here, we present high-resolution images of the putative inner lobes and new 
images of the overall radio structure of J0028$+$0035 resulting from our 
observations with LOw-Frequency ARray \citep[LOFAR,][]{vanhaarlem2013}, 
GMRT, and the Karl G. Jansky Very Large Array (VLA). 
Based on that material and the literature, we have collected radio flux densities in the range from 74\,MHz to 
14\,GHz for all components of J0028$+$0035. Our data base is characterized in 
Table\,\ref{table:radio_data}.

\subsection{LOFAR observations}\label{LOFAR_data}

Dedicated observations of J0028$+$0035 with LOFAR (project code: 
LC8\_007) were conducted in August 2017 (for details see 
Table\,\ref{table:radio_data}) in cooperation with the LOFAR Two-metre Sky 
Survey (LoTSS) Tier-1 Team -- see \cite{Shimwell2017} for the survey 
description. We observed the target in the HBA low-band Dual Inner mode 
using 24 core and 14 remote LOFAR stations. Total integration time was 
8\,h, including 10\% overhead, with one-second integration time, 244 
subbands with 16\,channels per subband in 110--190\,MHz filter with 200-MHz 
clock speed. To mitigate the influence of the ionosphere, we required 
the target's elevation to be over $30\degr$ above the horizon during the 
observations. Source 3C\,48 was used as a~flux density and bandpass 
calibrator in 10-min runs at the beginning and the end of the 
observations.

\begin{figure}
\includegraphics[width=1.0\linewidth]{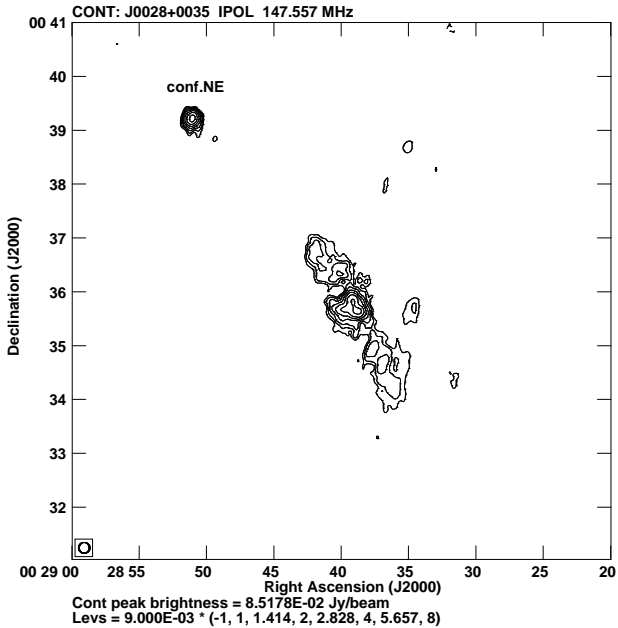}
\caption{LOFAR HBA image of J0028$+$0035 encompassing coincident 
conf.\,NE source.}
\label{fig:LOFAR}
\end{figure}

Standard reduction as well as flux density, phase, and 
bandpass calibration were provided by the Radio Observatory and the Tier-1 
Survey Team. Due to low observing frequency, large field of view of the 
instrument ($\sim$5$\degr$), and low declination of the target, we decided 
to increase the accuracy of the final map through the PreFactor 
pipeline,\footnote{https://github.com/lofar-astron/prefactor/} which 
prepares the data to any direction-dependent calibration and includes i.a. 
advanced flagging and removal of the initial clock offsets between the core 
and remote stations. Then, we applied the direction-dependent LOFAR Facet 
Calibration described in \cite{Weeren2016}. Both pipelines were run 
on CEP3 cluster allocated by the LOFAR Program Committee and 
the International LOFAR Telescope (ILT) director during the regular proposal 
evaluation stage. Apart from that, direction-dependent calibration and 
imaging were provided by the Tier-1 Survey Team with the use of 
ddf-pipeline.\footnote{https://github.com/mhardcastle/ddf-pipeline/} We were 
able to resolve the central region of J0028$+$0035 using Facet Calibration 
method whereas the extended, diffuse lobes were better 
calibrated and imaged with ddf-pipeline. The latter also allowed 
for better handling of interference from strong (above 1.3\,Jy at 
147.6\,MHz) northern source 4C\,$+$00.03. The resolutions of the maps 
attained after application of Facet Calibration and ddf-pipeline were
14.54\,$\times$\,6.25\,arcsec$^2$ and 12\,$\times$\,12\,arcsec$^2$, respectively, 
with the corresponding rms noise levels of 1 and 1.7\,mJy\,beam$^{-1}$.

The LOFAR image of the total structure of J0028$+$0035 is shown in 
Fig.\,\ref{fig:LOFAR} that also features a relatively strong coincident source 
towards north-east from the target source, afterwards referred to as 
conf.\,NE. The reference to the figure where the central structure of 
J0028$+$0035 observed with LOFAR is shown is given at the end of 
Section\,\ref{VLA_data}. The flux density measurements for different 
components of the target source and the conf.\,NE are shown in 
Table\,\ref{table:flux_densities}.

\begin{table*}
\caption{Observed flux densities of different components of J0028$+$0035}
\begin{tabular}{r c c c c c c c c c}
\hline
Frequency       &\multicolumn{8}{c}{Flux density (mJy)} & Reference\\
\multicolumn{1}{c}{(MHz)} & Total structure &  &\multicolumn{4}{c}{$<$--------------- Central structure -----------------$>$}& \multicolumn{2}{c}{Outer structure}&    \\  
                &                                   &  Conf. NE     & SW lobe     &   NE lobe      & Core       &  Blazar      &  SW lobe & NE lobe          &    \\
\multicolumn{1}{c}{(1)} & (2)                               &    (3)        &      (4)    &    (5)         &   (6)      &  (7)         &   (8)    & (9)              &(10)\\
\hline
         &                                    &               &             &                &            &              &          &              & 1\\
73.8     & $1593.0\pm288.0^{a}$               &               &             &                &            &              &          &              & 2\\
76    	 & $1581.6\pm158.2$                   & $301\pm97.0$  &             &                &            &              &          &              & 2\\  
84       & $1463.6\pm146.4$                   & $212\pm76.5$  &             &                &            &              &          &              & 2\\    
91.5     & $1297.6\pm129.8$                   & $144\pm66.7$  &             &                &            &              &          &              & 2\\    
99       & $1110.0\pm111.0$                   &               &             &                &            &              &          &              & 2\\    
107      & $1158.8\pm115.9$                   &               &             &                &            &              &          &              & 2\\    
114.5    & $992.5\pm99.3$                     & $203\pm47.7$  &             &                &            &              &          &              & 2\\      
122      & $970.4\pm97.0$                     &               &             &                &            &              &          &              & 2\\    
130      & $815.7\pm81.6$                     &               &             &                &            &              &          &              & 2\\    
143      & $811.3\pm81.1$                     &               &             &                &            &              &          &              & 2\\     
143.7    &                                    &               &             &                &            & $81\pm8$     &$218\pm33$& $250\pm38$   & p\\   
147.6    & $824\pm124$                        &               &$119^{c}\pm12$&$102\pm11$     &            &              &          &              & p\\
150.5    & $793.8\pm79.4$                     & $142\pm31.4$  &             &                &            &              &          &              & 2\\     
158      & $750.0\pm75.0$                     &               &             &                &            &              &          &              & 2\\    
166      & $732.3\pm73.2$                     &               &             &                &            &              &          &              & 2\\     
173.5    & $706.6\pm70.7$                     & $149\pm32.5$  &             &                &            &              &          &              & 2\\      
181      & $620.8\pm62.1$                     &               &             &                &            &              &          &              & 2\\     
189      & $579.5\pm58.0$                     &               &             &                &            &              &          &              & 2\\     
196.5    & $613.8\pm61.4$                     &               &             &                &            &              &          &              & 2\\      
200.5    & $585.5\pm58.6$                     & $130\pm26.4$  &             &                &            &              &          &              & 2\\      
204      & $568.6\pm56.9$                     &               &             &                &            &              &          &              & 2\\      
212      & $540.1\pm54.0$                     &               &             &                &            &              &          &              & 2\\     
219.5    & $553.5\pm55.4$                     & $99\pm23.7$   &             &                &            &              &          &              & 2\\       
227      & $512.9\pm51.3$                     &               &             &                &            &              &          &              & 2\\     
322.7    & $394.2\pm39.4$                     & $71.3\pm0.8$  &$59.2^{c}\pm6.4$&$48.9\pm4.9$ &            &$51.3\pm5.1$  &$91.9\pm9.2$&$101.0\pm10.1$&p\\
607.7    &                                    & $30.1\pm3.0$  &$38.0^{c}\pm4.0$&$27.5\pm3.0$ &            &$30.1\pm3.3$  &          &              &  p\\ 
607.7$^{b}$& $218.5\pm22.0$                   &               &$37.5^{c}\pm3.8$&$25.9\pm2.6$ &            &$35.4\pm3.6$  &$47.5\pm4.9$&$57.4\pm5.9$& p\\
1400.0   & $123.9\pm12.5$                     & $11.7\pm1.5$  &             &                &            &              &          &              & 3\\
1400.0   &                                    & $10.9\pm1.0$  &$22.0^{c}\pm2.2$& $16.5\pm1.7$&            &$20.3\pm2.0$  &$25.7\pm2.7^{d}$&$27.0\pm2.8^{d}$& 4\\
1437.4   &                                    &               &$22.8^{c}\pm2.3$&$18.4\pm1.8$ &            &$22.5\pm2.3$  &$28.5\pm3.0$&$24.3\pm2.5$& p\\
1519.4   &                                    &               &$22.9^{c}\pm2.3$&$17.9\pm1.8$ &$8.4\pm0.9$ &$21.3\pm2.2$  &          &              & p\\
3000.0   &                                    &               &$13.6^{c}\pm1.4$&$10.1\pm1.1$ &$4.9\pm0.5$ &$15.5\pm1.6$  &          &              & 5\\ 
4850.0   & $38\pm10$                          &               &             &                &            &              &          &              & 6\\
5469.0   &                                    &               &$8.5^{c}\pm0.9$&$6.7\pm0.7$   &            &$11.0\pm1.1$  &$4.9\pm0.5$&$5.1\pm0.5$  & p\\
5494.0   &                                    &               &$7.4^{c}\pm0.8$&$5.1\pm0.5$   &$2.7\pm0.3$ &$9.8\pm1.0$   &          &              & p\\
9016.8   &                                    &               &$5.8^{c}\pm0.6$&$4.6\pm0.5$   &            &$8.5\pm0.9$   &$2.1\pm0.2$& $2.3\pm0.2$ & p\\
14014.9  &                                    &               &$3.7^{c}\pm0.4$&$3.3\pm0.3$   &            &$6.1\pm0.6$   &          &              & p\\
\hline
\end{tabular}
\begin{flushleft}
{\em References} -- (1) \cite{Lane2014}; (2) \cite{Hurley-Walker2017}; (3) \cite{Condon1998}; (4) \cite{Becker1995}; (5) \cite{Lacy2020}; (6) \cite{Gregory1991}; (p) this paper. 
$^a$The original flux density of the VLSSr survey \citep[RBC scale;][]{Roger1973} was multiplied by a factor of 0.9 to suit the \cite{Baars1977} scale.
$^b$This has been taken from the tapered map.
$^c$This measurement includes also the core flux density.
$^d$This measurement is from the combined (with {\tt IMERG}) FIRST and NVSS map.\\
\end{flushleft}
\label{table:flux_densities}
\end{table*}

\begin{figure*}
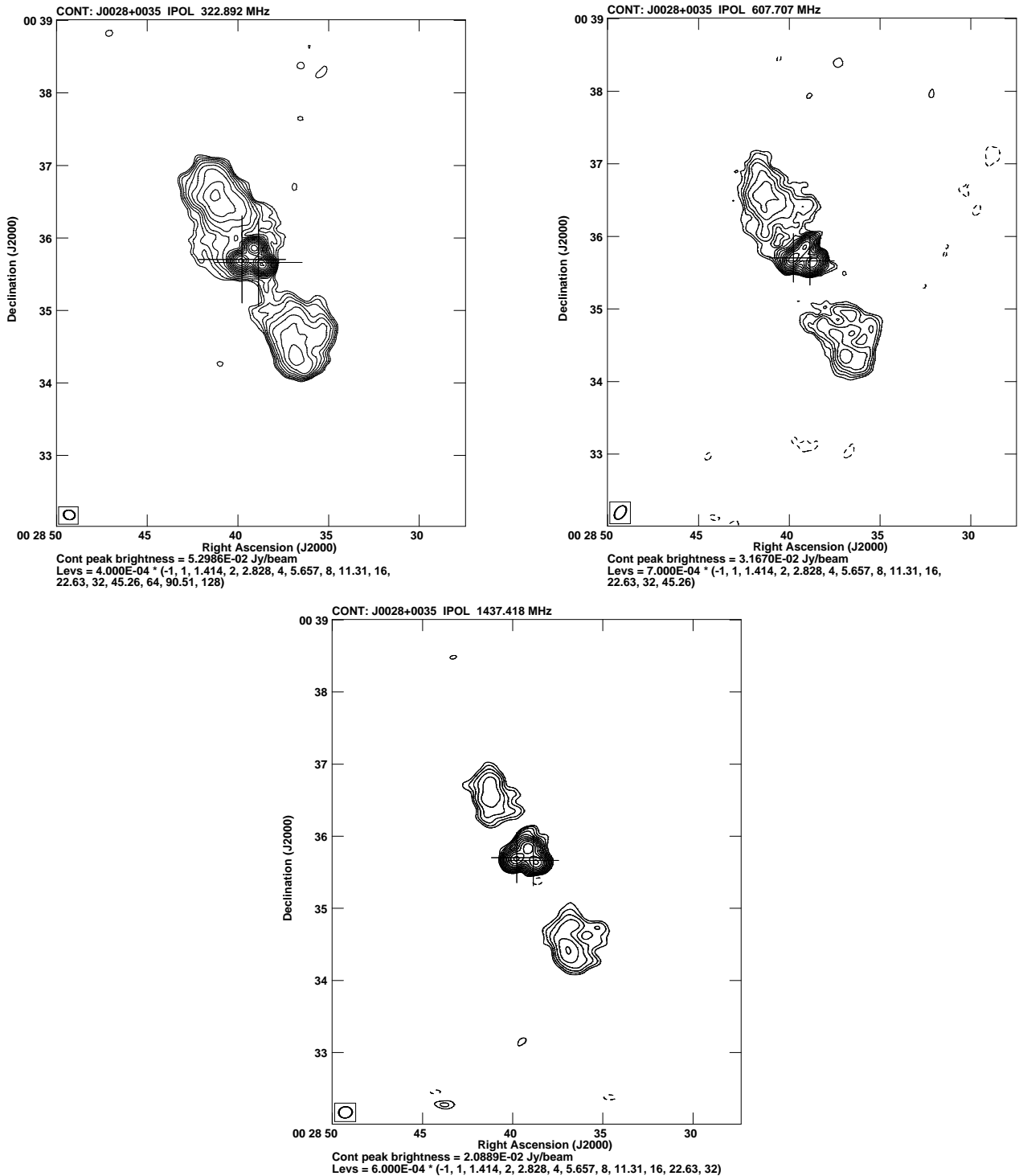

\includegraphics[width=0.45\linewidth]{J0028_PBAND.SUBIM.2.new.ps}
\hspace{1 cm}
\includegraphics[width=0.45\linewidth]{J0028_607MHZ.SUBIM.2.new.ps}
\includegraphics[width=0.45\linewidth]{J0028_GMRT.SUBIM.1.new.ps}
\caption{Images of J0028$+$0035 resulting from the GMRT observations at 
323\,MHz (upper left-hand panel), 608\,MHz (upper right-hand panel), and 1437\,MHz (bottom panel).
The optical position of the BL\,Lac object 5BZU\,J0028$+$0035 is marked with
the left-hand cross. The position of the SDSS\,J002838.86$+$003539.7 galaxy 
is marked with the right-hand cross.}
\label{fig:outer_GMRT}
\end{figure*}

\begin{figure*}
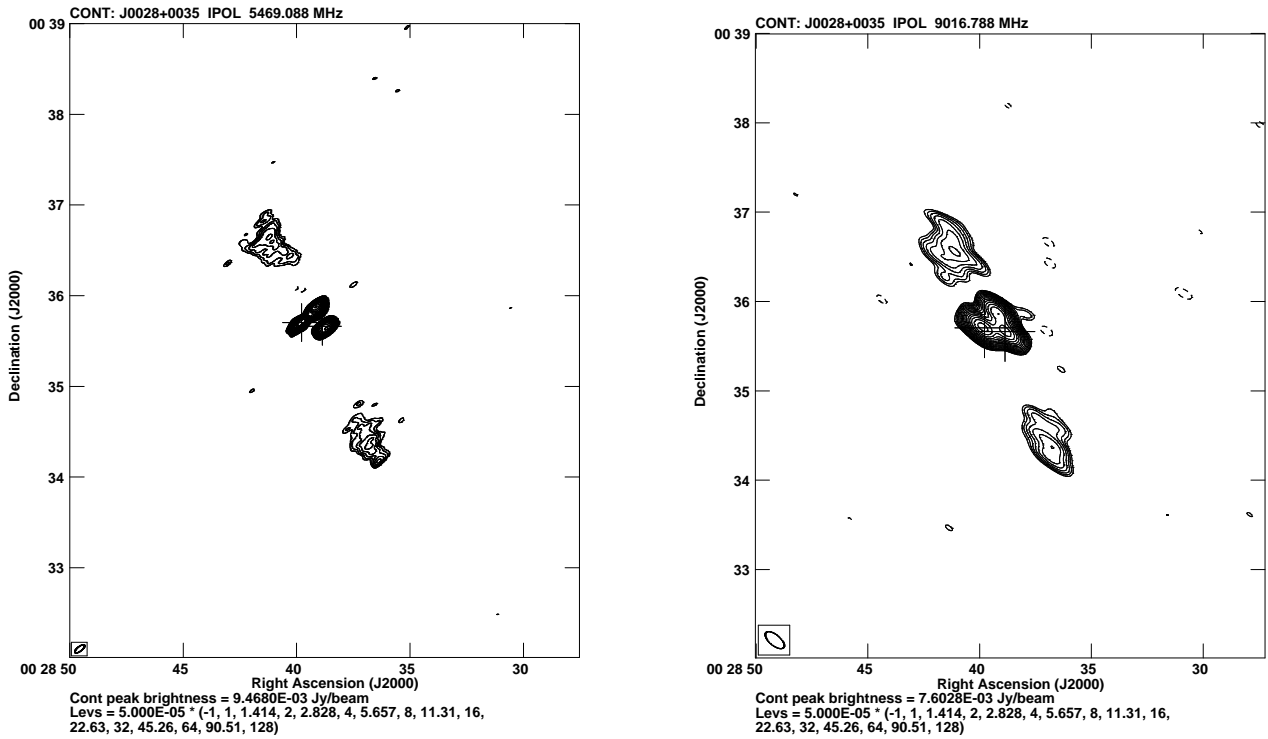

\includegraphics[width=0.45\linewidth]{VLA.C-band_outer.new.ps}
\hspace{1 cm}
\includegraphics[width=0.45\linewidth]{VLA.X-band.new.ps}
\caption{The VLA observations at 
5469\,MHz in C\,conf. (left-hand panel) and at 9017\,MHz in D\,conf. (right-hand panel).
The optical position of the BL\,Lac object 5BZU\,J0028$+$0035 is marked with 
the left-hand cross. The position of the SDSS\,J002838.86$+$003539.7 galaxy 
is marked with the right-hand cross.}
\label{fig:outer_VLA}
\end{figure*}

\begin{figure*}
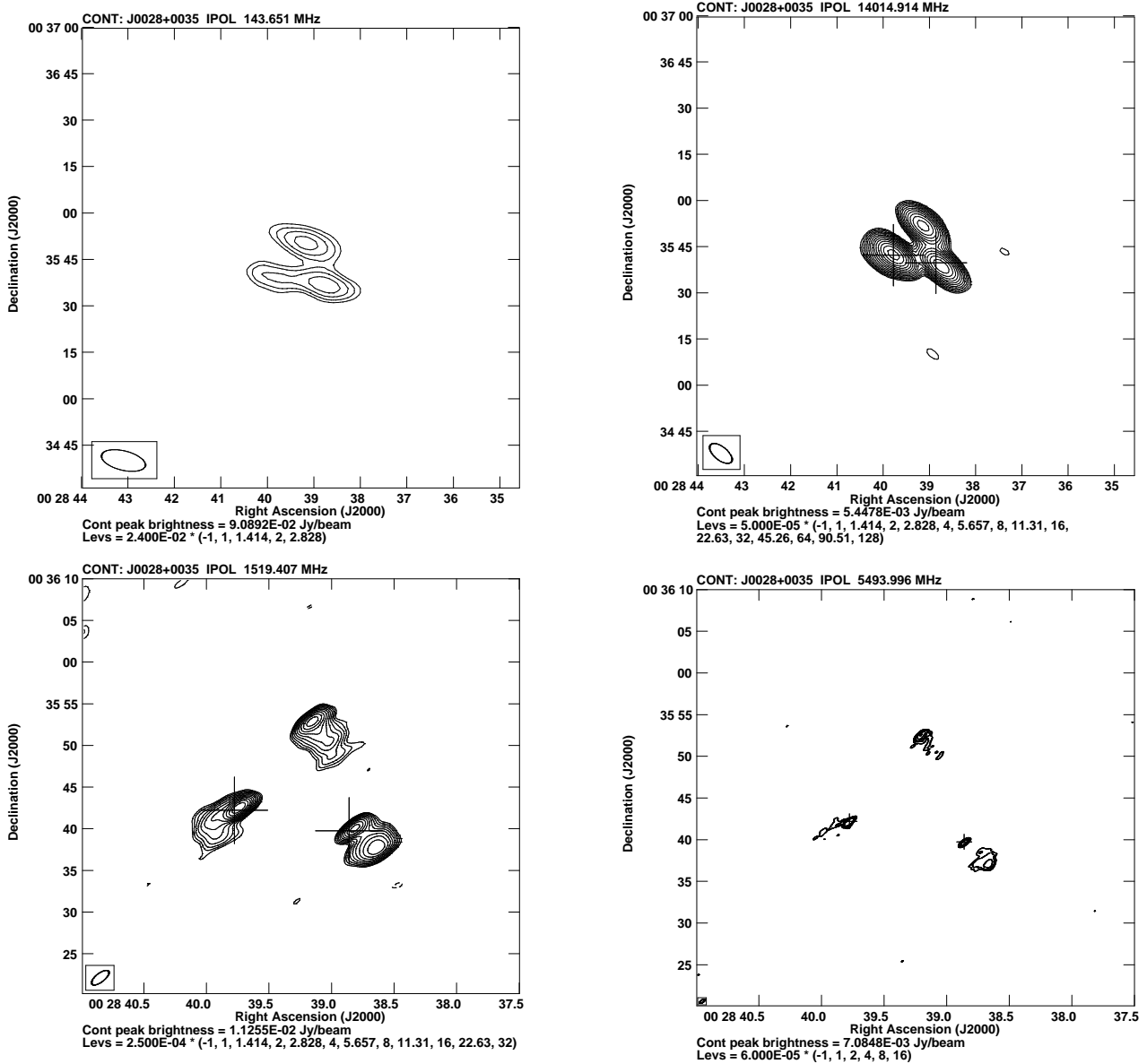

\includegraphics[width=0.45\linewidth]{J0028_LOFAR.INNER.new.ps}
\hspace{1 cm}
\includegraphics[width=0.45\linewidth]{VLA.K-band.new.ps}
\includegraphics[width=0.45\linewidth]{VLA.L-band.new.ps}
\hspace{1 cm}
\includegraphics[width=0.45\linewidth]{VLA.C-band.new.ps}
\caption{Central part of J0028$+$0035. Images resulting from LOFAR HBA
observations (upper left-hand panel) and the VLA observations at 14015\,MHz in D\,conf.
(upper right-hand panel), at 1519\,MHz in A\,conf. (lower left-hand panel), and at 5494\,MHz 
in A\,conf. (lower right-hand panel). The optical position of the BL\,Lac object 5BZU\,J0028$+$0035
is marked with the left-hand cross. The position of the
SDSS\,J002838.86$+$003539.7 galaxy is marked with the right-hand cross.} 
\label{fig:inner}
\end{figure*}

\subsection{GMRT observations}\label{GMRT_data}

We performed dedicated observations of J0028$+$0035 with GMRT in three 
frequency bands centred at: 323, 608, and 1437\,MHz. The project code was 32\_008.
The observations were conducted in September 2017 (for details see 
Table\,\ref{table:radio_data}). Data were recorded with 8-s integration 
time with the frequency band of 33.3\,MHz divided into 512 channels at 
323 and 608\,MHz or 256 channels at 1437\,MHz. The usual scheme of observing phase 
calibrator interlaced with the observation of the target source was adopted. 
Phase calibrator J0022$+$0014 (4C\,00.02) was used at each of the observed frequencies.
Flux density calibrator 3C\,48 was observed for about 15\,min at the beginning 
and at the end of observing at each frequency. The total integration time on the 
target source, which consisted of several 30-min exposures, was about 6.5, 2.5, 
and 5.5\,h at 323, 608, and 1437\,MHz, respectively.

The 323 and 608\,MHz data reduction was carried out following standard 
calibration and reduction procedures in {\sc Astronomical Image Processing
System} ({\sc aips}).\footnote{https://aips.nrao.edu/}
The data processing was automated with a pipeline based on the 
Source Peeling and Atmospheric Modeling (SPAM) package
\citep{IntemaPhD2009,Intema2014,Intema2009} that includes
direction-dependent calibration, modelling, and corrections for
dispersive phase delay which is mainly of ionospheric origin. Data 
were edited for strong radio frequency interference and then standard flux 
density, phase, and bandpass calibrations were applied to the source.
To produce deconvolved images, we used {\sc aips} {\tt IMAGR} task in which 
we subdivided the field of view covering the primary beam into a number of 
facets. To obtain an image that presents the diffuse outer lobes at 608\,MHz 
properly, we tapered the data at 20\,$k\uplambda$ so that the restoring 
beam is 12.49\,$\times$\,7.95\,arcsec$^2$. Several rounds of 
phase-based self-calibration were performed. The resultant image was then 
corrected for the primary beam using {\sc aips} {\tt PBCOR} task.

The 1437-MHz data were automatically flagged, reduced, and calibrated with 
package {\sc casa} ({\sc Common Astronomy Software
Applications}),\footnote{https://casa.nrao.edu/} using a pipeline originally developed by 
Russ Taylor in 2011 and modified by \cite{Ishwara-Chandra2020}. The 
resulting image was obtained using {\sc casa} task {\tt tclean} and its 
quality was improved using self-calibration. We tapered the original uv-data 
to extract the diffuse lobes with the restoring beam of 
11.17\,$\times$\,9.35\,arcsec$^2$. The image was also corrected for the 
primary beam.

The GMRT images are shown in Fig.\,\ref{fig:outer_GMRT}\footnote{For 
consistency, all its panels were plotted with {\sc aips} utility regardless of 
whether the images were generated in {\sc aips} or in {\sc casa}.}. Owing to 
tapering of 608 and 1437\,MHz ones, their resolution is comparable to 
that of the 323-MHz image (9.43\,$\times$\,7.79\,arcsec$^2$). The flux 
density measurements and their errors are shown in 
Table\,\ref{table:flux_densities}. The flux density calibration errors are 
assumed to be 10\% at all three frequencies.

\subsection{VLA observations}\label{VLA_data}

Five dedicated observations of J0028$+$0035 with the VLA in three 
configurations at four frequencies were conducted. They are listed in 
Table\,\ref{table:radio_data} in the lines denoted with `VLA-A',
`VLA-C', and `VLA-D' in column\,2. Standard continuum processing with 
VLA {\sc casa} Calibration Pipeline was carried out. Final images were obtained
with {\sc aips} task {\tt IMAGR}. Fig.\,\ref{fig:outer_VLA} presents the 
overall structure of J0028$+$0035 at 5469 and 9017\,MHz but with 
unresolved inner lobes whereas in Fig.\,\ref{fig:inner}, the inner 
lobes are shown in full detail while the outer lobes are 
not present either due to the `missing flux' effect caused by the resolution 
in A\,conf. (lower panels) or due to steepness of the spectrum of the outer 
lobes so that they are not visible at 14\,GHz (upper right-hand panel). The 
LOFAR image of the inner triple has been included as the fourth, upper left-hand 
panel of Fig.\,\ref{fig:inner}. The flux 
density measurements for different components of the target source and the 
errors of those measurements are shown in Table\,\ref{table:flux_densities}.

\subsection{Data extracted from surveys}
\label{Surveys_data}

In Table\,\ref{table:flux_densities}, we also display a number of data from the literature.
The lowest frequency at which we have data for our target is 74\,MHz. We used the \textit{VLA Low-Frequency Sky Survey} redux 
\citep[VLSSr;][]{Lane2012, Lane2014} that covers the sky north of declination $-30\degr$. The VLSSr represents a major improvement (e.g. correction of the ionospheric distortions and increase of dynamic range, as well as revision of the primary beam correction) 
to the original VLSS \citep{Cohen2007}. The VLSSr has a resolution of 75\,$\times$\,75\,arcsec$^2$ and an average 
map rms noise level of $\sim$0.1\,Jy\,beam$^{-1}$. The VLSSr maps and catalogue use the \cite{Roger1973} flux density scale.

The second public survey we used was the \textit{GaLactic and Extragalactic All-sky MWA} (GLEAM) 
survey carried out with the Murchison Wide-field Array \citep[MWA;][]{Lonsdale2009,Tingay2013}. 
GLEAM extends to the entire sky south of declination $+30\degr$ and is described in detail by \cite{Wayth2015}. The survey 
covers the frequency range between 72 and 231\,MHz with the bandwidth of 7.7\,MHz. The angular resolution of the
survey is $2\farcm5 \times 2\farcm2 \sec(\delta +26\fdg7)$ at 154\,MHz. The GLEAM data -- both the images and the
catalogue \citep{Hurley-Walker2017} -- are publicly accessible on the MWA Telescope
website.\footnote{http://www.mwatelescope.org/gleam}

J0028$+$0035 is well visible in the map of the TIFR GMRT Sky Survey 
(TGSS) First Alternative Data Release \citep[ADR1;][] {Intema2017} conducted 
at 147.5\,MHz. However, the TGSS~ADR1 data were compiled without the 
short-baseline visibilities within 0.2\,k$\uplambda$ from the centre of the 
(u,v)-plane of GMRT. For this reason, extended structures could not be 
properly imagined and can have underestimated flux densities. Indeed, the 
flux density of the whole structure of J0028$+$0035 extracted from the 
TGSS~ADR1 map is $540.5\pm56.4$\,mJy. This is much below the flux density 
measured at similar frequencies with GLEAM and LOFAR
-- see Table\,\ref{table:flux_densities}. Therefore, we did not 
use the TGSS~ADR1 data in our multifrequency analysis.

We also made use of three large VLA surveys: FIRST, \textit{NRAO VLA 
Sky Survey} \citep[NVSS;][]{Condon1998}, and \textit{VLA Sky Survey} (VLASS).
At 1400\,MHz, we used images extracted from FIRST and NVSS. Since about 60 per cent of the flux 
density is lost in the high-resolution FIRST maps, the FIRST and NVSS images 
were combined using {\sc aips} task {\tt IMERG}. The final merged maps were checked for flux density 
consistency by comparing the flux density of point sources in the FIRST and 
merged maps. The differences did not exceed 1 per cent. The 3-GHz data we used come from VLASS,
an all-sky (north of declination $-40\degr$) survey resulting from observations with the VLA in B-array configuration
conducted in a broad bandwidth of $2-4$\,GHz with the angular resolution 
$\sim$2.5\,arcsec and final sensitivity of 70\,$\mu$Jy beam$^{-1}$.
The VLASS observations began in September 2017 and will finish in 2024. The data of Stokes I, Q, U parameters
will be taken in three epochs to allow for the discovery of variable and transient radio sources. 
\cite{Lacy2020} presented the science case, the observational strategy for VLASS, and the results from early survey observations.
Early results in the form of Quick Look images of Stokes I, which use a~relatively simple imaging algorithm that results 
however in limits on the accuracy, are publicly available.\footnote{https://science.nrao.edu/vlass/vlass-data}

\subsection{Radio morphology of J0028$+$0035}

\begin{figure}
\includegraphics[width=\linewidth]{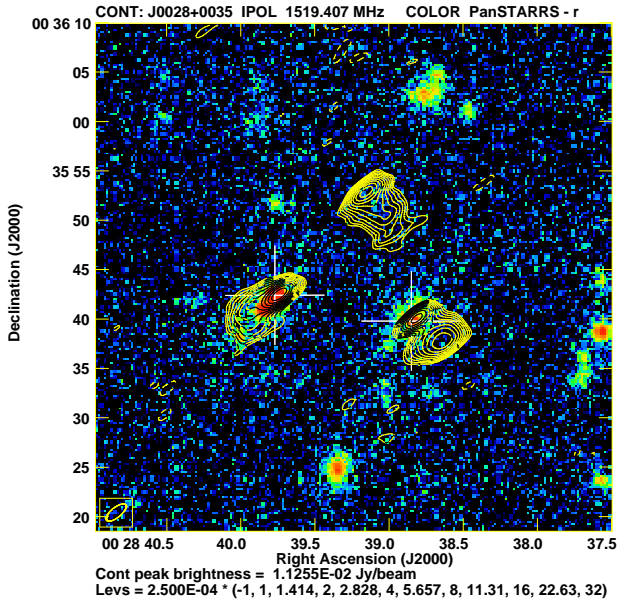}
\caption{The 1519\,MHz VLA A\,conf. image (contours) overlaid on PanSTARRS
$r$-band image.}
\label{fig:LbandVLAandPanSTARRS}
\end{figure}

The VLA A\,conf. images (Fig.\,\ref{fig:inner}, lower panels) unveil
the structure of the apparent inner triple. Additionally, we overlaid
the 1519\,MHz VLA A\,conf. image (contours) onto 
PanSTARRS\footnote{https://panstarrs.stsci.edu/} $r$-band image
\citep{Flewelling2020} -- see Fig.\,\ref{fig:LbandVLAandPanSTARRS}. If the
eastern component identified with 5BZU\,J0028$+$0035 is ignored, the 
remaining two are clearly an FR\,II-type mini-double hence J0028$+$0035 as 
a whole is, in fact, a DDRS. As seen in these images, 
5BZU\,J0028$+$0035 is of core-jet type. Given such morphological diversity 
of those components, the possibility that the inner triple of J0028+0035
is a result of gravitational lensing is unlikely. The feature identified with 
SDSS\,J002838.86$+$003539.7 appears as a separate object in the 5.5\,GHz 
A\,conf. image (Fig.\,\ref{fig:inner}, lower right-hand panel). Although it is 
located highly asymmetrically between the inner lobes, it might be
the core given that such extreme arm-length ratios are encountered in 
double-lobed radio sources like 3C\,254 \citep{Thomasson2006} or 3C\,459 
\citep{Thomasson2003}. It follows that SDSS\,J002838.86$+$003539.7 is the 
host galaxy of J0028$+$0035. Its spectroscopic redshift extracted from SDSS 
DR16 \citep{SDSSDR16} amounts to $z\!=\!0.398\,46\pm0.000\,08$. It must be noted
at this point that the SDSS DR16 spectroscopic redshift of 5BZU\,J0028$+$0035  
amounts to $z\!=\!0.686\,32\pm0.000\,25$. The self-evident discrepancy between these
two redshifts supports our inference that 5BZU\,J0028$+$0035 must be a coincidence.

The host galaxy redshift yields the scale of 5.33\,kpc\,arcsec$^{-1}$ thus 
the projected span of the outer lobes is 1093\,kpc while that of the inner 
lobes as measured in the 5.5\,GHz VLA A\,conf. image is 93\,kpc. 
These projected sizes translate to lobe lengths indicated in Table\,\ref{table:model_summary} for the assumed jet viewing angle.
It follows that J0028$+$0035 is a giant radio galaxy -- see 
\cite{Dabhade2020}, and references therein.

\begin{figure}
\includegraphics[width=1.0\linewidth]{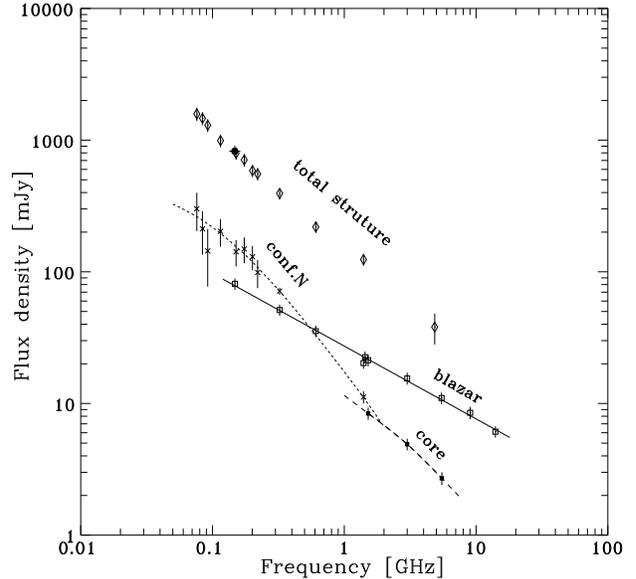}
\caption{Radio spectrum of the total structure of J0028$+$0035 and its core, 
as well as of the confusing source (conf.\,NE) located near the outer 
northern lobe and the blazar located in the vicinity of the central 
structure of J0028$+$0035}
\label{fig:spectr_obs}
\end{figure}

\begin{figure}
\includegraphics[width=1.0\linewidth]{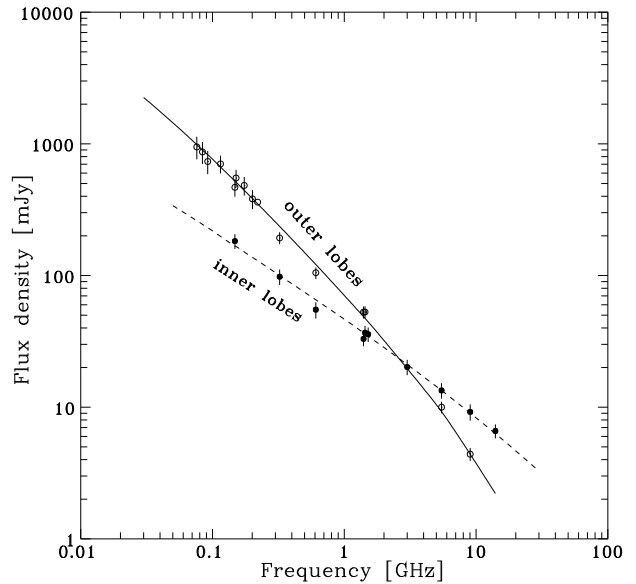}
\caption{{\sc dynage}-fitted spectra to the lobes: dashed line -- inner 
lobes, 
solid line -- outer lobes}
\label{fig:spectr_model}
\end{figure}

\subsection{Radio spectra of J0028$+$0035}

All the data itemized in Table\,\ref{table:flux_densities} have been used 
to construct radio spectra of individual components of the entire 
observed structure of J0028$+$0035, i.e. both pairs of its inner and outer
lobes, as well as two confusing sources: the blazar in the vicinity of the 
inner double and the one labelled conf.\,NE. The latter one
cannot be ignored since it is overlain by the MWA\,GLEAM beam size at the lowest
observing frequencies of 76, 84, and 92\,MHz. The resulting spectra of the core,
the blazar, and conf.\,NE are confronted in Fig.\,\ref{fig:spectr_obs} with the total structure data 
(column 2 in Table\,\ref{table:flux_densities}). It shows that the blazar spectrum is described by
a pure power-law function (solid line), while the spectrum of the conf.\,NE source
is strongly aged (dotted curve determined with a polynomial function 
$y=A+Bx+C\exp(\pm{x})$).

\section{Dynamical evolution analysis}
\label{dyn_evo}

\subsection{Numerical code}
\label{num_cod}

Similarly to \cite{Machalski2009,Machalski2010,Machalski2011,Marecki2016}, 
we carried out a dynamical analysis using {\sc dynage} code \citep{Machalski2007}.
It is a numerical implementation of an analytical model of the evolution of 
FRII-type radio source that combines the pure dynamical model of 
\cite{KaiserAlexander1997} and the \cite{Kaiser1997} (KDA) model of expected 
radio emission from a source under the influence of the energy loss processes.
For a given set of observables, the code allows for solving the inverse problem,
i.e. to determine four free parameters of the KDA model (see Table\,\ref{table:model_summary}):
(1) the initial power-law energy distribution of the relativistic electrons 
$p$ related to the effective injection spectral index $\alpha_{\rm inj}$,
(2) the jet power $Q_{\rm j}$, (3) the density of the external gaseous medium 
near the radio core $\rho_{0}$, and (4) the age of the source's radio 
structure $t$. This is obtained by fitting the model to the four 
observables, including the length and the volume of the lobes, and their 
radio spectrum, i.e. the slope and normalization at a number of observing 
frequencies.

\subsection{Application of the model and fitting procedure}

The analytical model described in Section\,\ref{num_cod} has been applied 
independently for the inner and outer lobes of J0028$+$0035.
The set of the model parameters with their values assumed for the given
pair of lobes is shown in Table\,\ref{table:model_summary}.

\subsubsection{Inner lobes}
\label{inn_lobes}
As a first step, we apply the model to the observational data of the inner lobes
because they are determined more precisely than those for the outer ones
(columns 4 and 5 vs. columns 8 and 9 of Table\,\ref{table:flux_densities}, respectively).
Although the spectra of 
both inner lobes have been derived, the flux densities of the SW lobe are confused by the
radio core at frequencies above 1.5\,GHz (column 4 in Table\,\ref{table:flux_densities}). While 
extraction of the core's spectrum between 1.5 and 5.5\,GHz (column 6 in
Table\,\ref{table:flux_densities}) has been successful, its extension towards lower frequencies is problematic.
An effort made to estimate the core excess in flux densities in column (4) of
Table\,\ref{table:flux_densities}, aroused suspicion of a high symmetry in the luminosities of both inner
lobes. Unlike in the case of DDRS\,J1706$+$4340 \citep{Marecki2016} where luminosities of the inner lobes are
highly asymmetric, the inner lobes of J0028$+$0035 show unusually asymmetric
separation from the core. It is obvious that such asymmetries induce significantly
different values of the model parameters for individual lobes. We thus apply
the modelling procedure to the whole inner structure setting up its spectrum as a double 
of the flux densities in column (5) of Table\,\ref{table:flux_densities} and using
one half of the sum of the opposite lobes' lengths.

Following \cite{Marecki2016}, we investigate two alternative models. In the first one, a power-law 
density distribution of the external gaseous medium with a standard exponent $\beta=1.5$ is assumed, 
while in the other one we assume that the inner lobes evolve into an almost uniform medium with 
$\beta=0.1$ inside a cocoon formed and inflated by the material of the primary jet flow. 
We also assume that the lobes are filled with magnetic fields and relativistic particles 
governed by a relativistic equation of state $\Gamma_{l}=4/3$.
The best-fitting models results are shown in columns (3) and (4) of Table\,\ref{table:in_out_densities} (predicted flux
densities) and in columns (2) and (3) of Table\,\ref{table:model_deriv} (predicted values of the main model and
derivative physical parameters of the inner lobes' structure). The resulting
best-fitted spectrum of these lobes -- which is almost the same in 
both models either with $\beta=1.5$ or with $\beta=0.1$) 
-- is shown in Fig.\,\ref{fig:spectr_model} (dashed curve).

\subsubsection{Outer lobes}
\label{out_lobes}

As shown in Table\,\ref{table:flux_densities}, the flux densities of the pair of outer lobes are
determined at the observing frequency range from 323\,MHz up to 9\,GHz and
supported with the data at the LOFAR\,HBA frequencies. However, the GLEAM
survey data enable supplementing the spectrum of those lobes in the range 
between 76 and 220\,MHz. In order to carry out such a procedure, especially at the
lowest frequencies range, we subtract the sum of the inner lobes model fit 
(column 2 of Table\,\ref{table:fitted_densities}; also cf. columns 3 and 4 of Table\,\ref{table:in_out_densities}), 
the blazar extrapolated power-law fit (column 3 of Table\,\ref{table:fitted_densities}), and the polynomial-form fit to the GLEAM
data (column 4 of Table\,\ref{table:fitted_densities}) from the flux densities in column 2 of Table\,\ref{table:flux_densities}.

Likewise to the inner structure, the best-fitting-model-predict\-ed flux densities are 
shown in column 6 of Table\,\ref{table:in_out_densities} and the model and derivative parameters are shown
in column 4 of Table\,\ref{table:model_deriv}. Similarly, the resulting spectrum of the outer lobes
is shown in Fig.\,\ref{fig:spectr_model} (solid curve).

\begin{table*}
\caption{Summary of the model parameters}
\begin{tabular}{lccc}
\hline
Parameter    & Symbol   & Inner lobes    & Outer lobes \\
\hspace{5mm}(1) & (2)  & (3)  & (4)   \\
\hline
{\em Observed}\\
Angular size of lobe ($\times 2$)           & $LAS$               & 17.4 arcsec & 205 arcsec \\
Length of lobe ($\times 2$)                 & $D$                 & 99 kpc      & 1163 kpc \\
Aspect ratio of lobe                        & $R_{\rm T}$         & 3.5         & 2.6   \\
Radio spectrum; i.e. monochromatic          & $\alpha_{\rm\nu}$ \\
\hspace{5mm}luminosity at a number of observing \\
\hspace{5mm}frequencies; $i=1,2,3$...       & $P_{\rm\nu,i}$      & Note(1)      & Note(2) \\
\vspace{1mm}
{\em Set}\\
Adiabatic index of jet material             & $\Gamma_{\rm j}$    & 4/3          & 4/3 \\
Adiabatic index of lobe material            & $\Gamma_{\rm\ell}$  & 4/3          & 5/3 \\
Adiabatic index of ambient medium           & $\Gamma_{\rm x}$    & 5/3          & 5/3 \\
Adiabatic index of magnetic field           & $\Gamma_{\rm B}$    & 4/3          & 4/3 \\
Minimum of initial electron Lorentz factor  & $\gamma_{\rm min}$  & 1            & 1   \\
Maximum of initial electron Lorentz factor  & $\gamma_{\rm max}$  & 10$^{7}$     & 10$^{7}$ \\
Core radius of ambient density distribution & $a_0$               & 2 kpc        & 2 kpc \\
Exponent of ambient density distribution    & $\beta$             & 1.5, 0.1     & 1.5 \\
Jet viewing angle                           & $\Theta$            & 70${\degr}$  & 70${\degr}$ \\
\vspace{1mm}
{\em Free}\\
Jet power                                   & $Q_{\rm j}$(W)  &\\
External density at core radius             & $\rho_0$(kg\,m$^{-3}$)   &\\
Exponent of initial power-law energy \\
\hspace{5mm}distribution of relativistic electrons  & $p$   &$=1+2\alpha_{\rm inj}$ \\
Source (lobe) age                           & $t$(Myr) &\\
\hline
\end{tabular}

\vspace{3mm}

(1) and (2) -- relevant luminosities are calculated with observed flux densities 
shown in columns 3 and 5 of Table\,\ref{table:in_out_densities}, respectively.
\label{table:model_summary}
\end{table*}

\begin{table*}
\caption{Model fit predicted or extrapolated flux densities (in mJy) of different 
components of the total structure of J0028$+$0035 used to extract 76--220\,MHz flux 
densities of its outer lobes from the GLEAM data base (shown in column\,2 of Table\,\ref{table:flux_densities})}
\begin{tabular}{rcccccc}
\hline
\multicolumn{1}{c}{Frequency}      & Inner lobes        &Blazar        & Conf. NE      & & Sum of column & Outer lobes \\
MHz           & model fit          & lin.extrap.   & exp.func.fit. & & (2)+(3)+(4) & GLEAM data--Sum\\
\multicolumn{1}{c}{(1)}        &    (2)             &    (3)       &    (4)      & & (5)        &  (6) \\ 
\hline
           &            	&              &             & &        &   \\
76         & 259.0            	& 113.3        & 262.3       & & 634.6  & 947.0 \\
84         & 242.7           	& 107.2        & 245.8       & & 595.7  & 867.9 \\
91.5       & 229.6           	& 102.2        & 230.8       & & 562.6  & 735.0  \\
114.5      & 198.5           	& 90.3         & --          & & 288.8  & 703.7   \\
150.5      & 166.0          	& 77.6         & --          & & 243.6  & 550.2 \\
173.5      & 151.2          	& 71.7         & --          & & 222.9  & 483.7\\
200.5      & 137.5           	& 66.2         & --          & & 203.7  & 381.8  \\
219.5      & 129.5           	& 63.0         & --          & & 192.5  & 361.0\\

\hline
\end{tabular}
\label{table:fitted_densities}
\end{table*}

\begin{table*}
\caption{Observed (or extracted) flux densities and the best-fitting flux
densities (in mJy) for the inner and outer lobes of J0028$+$0035}
\begin{tabular}{rccccccccc}
\hline
\multicolumn{1}{c}{Frequency} & & & Inner lobes    &\multicolumn{2}{c}Model fit & & &Outer lobes & Model fit \\
\multicolumn{1}{c}{(MHz)} & & &                    & $\beta=1.5$     &$\beta=0.1$ \\
\multicolumn{1}{c}{(1)}   & & &    (2)             &    (3)          & (4)        & & &  (5)      & (6)        \\ 
\hline
           & & &             	&                   & & &                              &            \\
76         & & &          	        & 259.0$^a$ & & & &  947$\pm$184$^{b}$          & 985.7    \\
84         & & &          	        & 242.7$^a$ & & & &  868$\pm$165$^{b}$          & 897.2    \\
91.5       & & &           	        & 229.6$^a$ & & & &   735$\pm$147$^{b}$         & 827.3    \\
114.5      & & &           	        & 198.5$^a$ & & & &  704$\pm$109$^{b}$          & 667.5    \\
147.6      & & & 182.5$\pm$23           & 167.8     & 165.4 & & &  468$\pm$71           & 519.3    \\
150.5      & & &          	        & 166.0$^a$ & & & &  550$\pm$85$^{b}$           & 510.7    \\
173.5      & & &           	        & 151.2$^a$ & & & &  483$\pm$78$^{b}$           & 443.6    \\
200.5      & & &           	        & 137.5$^a$ & & & &  382$\pm$63$^{b}$           & 383.7    \\
219.5      & & &                        & 129.5$^a$ & & & &  361$\pm$60$^{b}$         & 350.2    \\
322.7      & & & 97.8$\pm$13.1          & 100.3     & 99.3 & & &  193$\pm$20            & 236.1    \\
607.7      & & & 55.0$\pm$7.8           &  65.4     & 65.3 & & &  105$\pm$11            & 121.3    \\
1400.0     & & & 33.0$\pm$4.0           &  36.7     & 37.0 & & &  52.7$\pm$5.5          & 48.7     \\
1437.4     & & & 36.8$\pm$4.6           &  36.0     & 36.4 & & &  52.8$\pm$5.5          & 47.3     \\
1519.4     & & & 35.8$\pm$4.5           &  34.6     & 35.0 & &   \\
3000.0     & & & 20.2$\pm$2.7           &  21.1     & 21.4 & &   \\
5469.1     & & & 13.4$\pm$1.8           &  13.4     & 13.6 & & &  10$\pm$1              & 9.3      \\
9016.8     & & & 9.2$\pm$1.3            &   9.0     & 8.9  & & &  4.4$\pm$0.4           & 4.4      \\
14014.9    & & & 6.6$\pm$0.8            &   6.3     & 6.1  & &                          &          \\
\\                                  
$\chi^{2}_{\rm red}$&               & & & 0.494             & 0.572& & & & 0.805\\
\hline
\end{tabular}
\begin{flushleft}
$^a$The model predicted flux densities shown in column 2 of Table\,\ref{table:fitted_densities}. \\
$^b$The flux densities extracted from the GLEAM data base (cf. column 6 in Table\,\ref{table:fitted_densities}).
\end{flushleft}
\label{table:in_out_densities}
\end{table*}

\begin{table}
\caption{Model and derivative parameters for the inner and the outer 
structure (lobes) of J0028$+$0035}
\begin{tabular}{lcccc}
\hline
Parameters                &\multicolumn{2}{c}{Inner lobes} & & Outer lobes \\ 
                          &$\beta=1.5$ & $\beta=0.1$& &\\
\hspace{5mm}(1)           &  (2)       &  (3)  & & (4)    \\
\hline
{\em Model}\\
$\alpha_{\rm inj}$          & 0.63&0.64 &   & 0.59  \\
$Q_{\rm j}(\times 10^{37}$W)& 2.17&2.29 &   & 2.20  \\
$\rho_{0}(\times 10^{-22}$kg\,m$^{-3}$) & 0.081  &0.0048 &   & 2.67  \\
$t$(Myr)                    & --& -- &  &  245  \\
$t_{\rm j}$(Myr)            & 3.2&3.6 &   &  234  \\
\\
{\em Derivative}\\
$v_{\rm h}/c(\times 10^{-3})$               & 43.2  & 27.4   &   & 6.77  \\
$\rho_{(D/2)}(\times 10^{-24}$kg\,m$^{-3}$) & 0.066 & 0.3475 &   & 0.054 \\
$p_{\rm\ell}(\times 10^{-13}$N\,m$^{-2}$)   & 7.60  & 6.78   &   & 0.26  \\
$B_{\rm\ell}$(nT)                           & 1.61  & 1.52   &   & 0.28  \\

\hline

\end{tabular}
\label{table:model_deriv}
\end{table}

\begin{table*}
\begin{center}
\caption{Properties of J0028$+$0035 and selected DDRSs}

\begin{tabular}{llllllrrlllll}
\hline
Source       &  $z$     & $\log$M$_{\rm BH}^c$ & l$_{\rm inn}^c$ & l$_{\rm out}^c$ & $\alpha_{\rm inj}^{\rm inn}$& $\alpha_{\rm inj}^{\rm out}$& $t_{\rm in}$  & $t_{\rm out }$ &   $t_{\rm quies}$ & Q$_{\rm j}^{\rm inn}$      & Q$_{\rm j}^{\rm out}$       &   Reference \\
J2000 name   &  	&  (M$_{\odot}$) &  (kpc)       &  (kpc)      &                    &                     & (Myr)          &   (Myr)        &    (Myr)          & ($\times10^{37}$ W)& ($\times10^{37}$ W) &      \\
  (1)        &  (2)     &    (3)         &   (4)        &   (5)       &      (6)           &  (7)                &  (8)           &   (9)          &     (10)          & (11)               & (12)                &  (13) \\
\hline  
J0028$+$0035          &  0.3985            &    8.16        & 93           & 1093        &0.64               & 0.59           & 3.6            &245             & 11                &  2.29            & 2.20              &  p      \\    
J0041$+$3224          &  0.45$^b$          &    --          & 172          & 974         &0.60 		     & 0.62    	      & 4.0            &105             & 11                &  2.49            & 14.4              &  1      \\
J0116$-$4722$^a$      &  0.1461            &    --          & 445          & 1441        &$0.70$             &  $0.62$        & 1.00$-$28.00   &66$-$236        &  1.4$-$65.4       &   --             & --                &  2, 7   \\
J0840$+$2949$^a$      &  0.0647            &    8.32        & 39           & 533         &$0.83$             &  $0.81$        & 0.12$-$33.00   &$>200$          &  2.0$-$102.0      &   --             & --                &  3, 7   \\
J1158$+$2621$^a$      &  0.1121            &    7.96        & 139          & 484         &$0.77$  	     &  $0.79$        & 0.50$-$4.90    &113             &  6.6$-$11.0       &   --             & 1.7               &  2, 7   \\
J1352$+$3126          &  0.045             &    8.15        & 1            & 180         &0.59  	     &  0.61   	      & 0.3            &62              &  0.7              &  0.21            & 0.21              &  4      \\
J1453$+$3308          &  0.2482            &    9.02        & 159          & 1299        &0.66               &  0.50  	      & 5.0            &104             & 24                &  0.32            & 3.92              &  1      \\
J1548$-$3216          &  0.1082            &    --          & 312          & 962         &0.61  	     &  0.51   	      & 9.2            &132             & 30                &  0.14            & 1.59              &  1      \\
J1706$+$4340          &  0.525$^b$         &    --          & 194          & 687         &0.55               &  0.53          & 12             &260$-$300       & 27                &  2.63            & 2.59              &  5      \\   
J1835$+$6204$^a$      &  0.5194            &    --          & 372          & 1378        &$0.86$             &  $0.82$        & 1.34$-$2.25    & 22             & 1.0$-$6.6         &                  & 28.5              &  6, 7   \\
\hline
\end{tabular}
\begin{flushleft}

{\em References} -- p: this paper, 1: \cite{Machalskietal2011}, 2: \cite{Konar2013}, 3: \cite{Jamrozy2007}, 4: \cite{Machalski2016}, 
5: \cite{Marecki2016}, 6: \cite{Konar2012}, 7: \cite{KonarHardcastle2013}.
Column description is as follows: Column 1: J2000 name of the source, 
Column 2: redshift of the source, Column 3: black hole mass, Columns 4 and 5: size of the inner and outer double, respectively, Columns 6 and 7: 
alpha injection of the inner and outer double, respectively, Columns 8 and 9: age of the inner and outer double, respectively, Column 10: duration of the 
quiescent phase, Columns 11 and 12: jet power of the inner and outer double, respectively, Column 13: reference to the spectral age and other parameters of the source.
$^a$For this source the data have been taken from spectral modelling rather than from dynamical modelling as for other sources.
$^b$Photometric redshift.
$^c$Data taken from or calculated according to \cite{Kuzmicz2017}.

\end{flushleft}
\end{center}
\label{table:grgs_comparison}
\end{table*}

\section{Discussion}
\label{discussion}

\subsection{General remarks}
\label{discusion:general}

\begin{enumerate}

\item J0028$+$0035 is one of the most distant DDRSs known -- see table\,1 in
\cite{Kuzmicz2017}.

\item For frequencies lower than 3\,GHz, the total flux density of the outer 
lobes of J0028$+$0035 is larger than that of its inner ones. This trend is 
reversed for frequencies larger than about 3\,GHz.

\item For comparison with J0028$+$0035, some characteristic parameters of 
nine other well-studied DDRSs are presented in Table\,7. 
Noteworthy is the fact that the age of the outer structure of J0028$+$0035 
is one of the largest in this sample, the jet powers for both epochs 
of activity are similar, and the values of $\alpha_{\rm inj}$ are 
similar as well.
This has already been noticed for DDRSs -- see e.g. \cite{KonarHardcastle2013}.

\item J0028$+$0035 has one of the smallest -- 0.085 -- inner to outer lobes 
size ratio among giant DDRSs from \cite{Kuzmicz2017} sample. 
This is a clear hint that the inner double must be young -- see Sect.\,\ref{discusion:age}.

\end{enumerate}

\subsection{Asymmetries of the radio structure of J0028$+$0035}
\label{discusion:assymetry}

We already mentioned the asymmetry of the inner structure in 
Sect.\,\ref{inn_lobes} but this is not the only one in this object. We 
estimated the bending angle which is the complement of the angle between the 
lines connecting the brightest points in the lobes with the core. The 
bending angle is about $11\degr$ and $27\degr$ for the outer and inner lobes, 
respectively. We also estimated the arm-length ratio, $q$, i.e. the ratio of 
distances between the core and the hotspots or -- in the case of the outer 
lobes that are devoid of hotspots -- maxima of radio emission. To find 
those estimates, we used measurements carried out at 323\,MHz 
(Fig.\,\ref{fig:outer_GMRT}, upper left-hand panel) and at 5494\,MHz 
(Fig.\,\ref{fig:inner}, lower right-hand panel). The arm-length-ratio of the 
outer lobes is $q\!\simeq\!1.27$, the southern lobe being the longer whereas 
the arm-length ratio of the inner lobes is $q\!\simeq\!4.33$ and here the 
northern lobe is the longer one. Such a large asymmetry of the inner lobes 
of a~DDRS is atypical as they are usually more symmetric than the outer 
ones \citep{Saikia2006, Nandi2012}.

According to the standard interpretation, asymmetry of lengths 
between the two opposite lobes is a light-traveltime effect \citep[see 
e.g.,][]{Longair1979}. The size ratio between the longer and shorter lobe 
amounts to $$q=[1+(v_{\rm h}/c) \cos\theta]/[1-(v_{\rm h}/c) \cos\theta],$$ 
which yields the advance speed of the jets $$v_{\rm h} = 
(q-1)/(q+1)(c/\cos\theta).$$ From the values of $q$ determined 
above along with the condition $\mid\cos\theta\mid \leq\!1$, we get $v_{\rm 
h}/c\geq$ 0.1189 and $v_{\rm h}/c\geq$ 0.6248 for the outer and inner lobes, 
respectively. (Such a dominance of the latter velocity is typical for
DDRSs -- see e.g. \cite{Kaiser2000, Schoenmakers2000b, Machalski2011}.)
On the other hand, the values of the expansion velocity from 
the {\sc dynage} modelling -- see Table\,\ref{table:model_deriv} -- are over an 
order of magnitude smaller. It seems, therefore, that the lobe length 
asymmetry is not a pure orientation effect. Instead, the inhomogeneity
of the intergalactic medium (IGM) may play a role here. Another possibility that could also be 
taken into account is that the changes of the bending angle of the external 
and internal structure may indicate that the jet changes its orientation 
between the two epochs of activity. A~change of the direction of 
the inner and outer structures' jets is not exceptional in DDRSs. In 
Section\,\ref{intro}, we already mentioned J2333$-$2343 but also 
the morphologies of J1352$+$3126 \citep[3C\,293;][]{Bridle1981, Akujor1996, 
Beswick2004, Joshi2011, Machalski2016}, J0709$-$3601 
\citep{Subrahmanyan1996, Saripalli2013}, and J1328$+$2752 \citep{Nandi2017} 
show clear evidence for the restart of the jets' activity accompanied by 
their substantial axis change. Moreover, GRGS\,J0009$+$1244 
\citep[4C\,12.03;][] {Leahy1991} and GRGS\,J1513$+$2607 
\citep[3C\,315;][]{Saripalli2009} have large-scale X-shaped morphologies 
which 
may also be caused by that.

\subsection{Density, pressure, and kinetic temperature}

J0028$+$0035 lacks X-ray observations, which makes direct estimation of the 
parameters of the IGM surrounding this DDRS -- i.e. 
density, pressure, and kinetic temperature -- impossible. Nevertheless, 
crude estimates of the ambient kinetic temperature and the sound speed in 
the ambient gaseous environment are possible using the physical parameters 
shown in Table\,\ref{table:model_deriv}. From the equation of state for a 
perfect gas, $p=nkT$, one can find that $$kT=\mu m_{\rm H}p/\rho\sim\mu 
m_{\rm H}p_{l}/\rho_{(D/2)},$$ where $n=\rho/\mu m_{\rm H}$ and 
$p_{l}$ is the pressure inside the lobe. 
The value of the ambient density at the end of the lobe, $\rho_{(D/2)}$, 
along with other necessary parameters' values are shown in Table\,\ref{table:model_deriv}.
Assuming the mean molecular weight for ionized gas 
$\mu\!=\!0.62$, we find for the outer lobes $kT\!=\!3.1$\,keV. It follows 
that the sound speed $c_{\rm s}\!=\![(\Gamma_{x}kT)/(\mu m_{\rm H})]$ for the 
outer lobes amounts to 0.00298\,$c$ and the ratio $v_{\rm h}/c_{\rm 
s}\!=\!2.27$. This means that the longitudinal expansion of these lobes is 
still supersonic. Although no hotspots can be seen in the outer lobes, 
traces of bow shocks are visible in the low-frequency radio maps -- see the
upper panel of Fig.\,\ref{fig:outer_GMRT}. The compatibility of the above 
ambient temperature with the X-ray temperatures in samples of powerful radio 
sources \citep{Belsole2007} is worth noting.

To estimate the kinetic temperature inside the cocoon surrounding the 
inner lobes, whose gaseous environment is likely to be 
modified by the jet flow during the earlier phase of activity, 
we apply the model of $\beta=0.1$ that corresponds to a rather smooth medium.
We find $kT\!=\!12.6$\,keV and $c_{\rm s}\!=\!0.00612\,c$. Dividing the 
corresponding values of the advance speed of the inner lobes $v\rm_{h}$ (see 
Table\,\ref{table:model_deriv}) by the relevant sound speed obtained above, 
we find the Mach number of 4.56. This value is twice as large as that for the outer 
lobes and the obtained kinetic temperature is four times higher. 
Meanwhile, the overall temperature of the very luminous cluster of galaxies 
RX\,J1347.5$-$1145 is $kT\sim$10\,keV \citep{Allen2002,Gitti2004}. 
The high value of $kT$ of the inner lobes of J0028$+$0035 is, however, comparable to the one 
\cite{Birzan2017} found for some high-redshift AGNs. Although the model with $\beta=1.5$ has
a lower value of $\chi^{2}_{\rm red}$ (see the last row of Table\,\ref{table:in_out_densities}), 
it leads to a prohibitively high value of kinetic temperature for the inner lobes thus it has to be rejected.

\subsection{Age and intermittent activity}
\label{discusion:age}

The age solutions for the inner and outer lobes collected in 
Table\,\ref{table:model_deriv} suggest that the last renewal of the jet 
activity took place about 3.6\,Myr ago hence its present age is about 
1.3\,per cent of the age estimated for the outer lobes' structure. On the 
other hand, the length of the period of quiescence emerging from the model fits, 
$t-t_{\rm j}$, is of about 11\,Myr. These values, when compared with 
those for other DDRSs listed in Table\,7, appear as typical. 
However, the ratio of inner to outer lobes' age for J0028$+$0035 is one of 
the lowest among those objects. (The only clear exception is J1352$+$3126.)

Assuming similar dynamical and spectral age of the inner structure and 
taking into account the dynamically estimated age of 3.6\,Myr, and magnetic 
field strengths of 1.52\,nT we calculated the break frequency of the 
synchrotron emission to be of about 32\,GHz. It would be possible to prove 
the above estimation with sensitive VLA observations at {\em K}\,band ($18-26.5$\,GHz), 
{\em Ka}\,band ($26.5-40$\,GHz) and {\em Q}\,band ($40-50$\,GHz) and/or, in the future, 
with the Atacama Large Millimeter/submillimeter Array using band\,1 
($35-50$\,GHz) receiver.

\section{Summary}
\label{summary}

\begin{enumerate}

\item We conducted an observing campaign targeted at J0028$+$0035, a giant 
radio source that has atypical morphology: a triple straddled by a pair of 
relic lobes (Fig.\,\ref{fig:FIRST}). It was observed with LOFAR, GMRT, and the VLA at a variety of 
frequencies (Table\,\ref{table:radio_data}). We present the images resulting 
from those observations (Figs\,\ref{fig:LOFAR}--\ref{fig:inner}). The 
conclusion we draw from that observational material is that the inner 
triple consists of an FR\,II-type double pertinent to J0028$+$0035 and a coincident 
source 5BZU\,J0028$+$0035 -- a BL\,Lac object whose redshift 
($z\!=\!0.686\,32$) does not correspond with that of the host galaxy of 
J0028$+$0035 ($z\!=\!0.398\,46$). Therefore, J0028$+$0035 as a whole is 
a~double--double source.

\item We supplemented the flux density measurements made using our data with 
those extracted from a number of radio sky surveys 
(Table\,\ref{table:flux_densities}). With that data base covering a very wide 
frequency range from 74\,MHz to 14\,GHz, we constructed the radio spectra 
(Fig.\,\ref{fig:spectr_obs}) of different parts of J0028$+$0035. We modelled 
those spectra with {\sc dynage} code (Fig.\,\ref{fig:spectr_model}). 
This way we were able to determine physical parameters for both the outer and
the inner lobes (Table\,\ref{table:model_deriv}).

\item We compared the values of some of those parameters derived for 
J0028$+$0035 with the values of respective parameters for a number of DDRS 
(Table\,7). It appears that J0028$+$0035 is a 
well-behaved DDRS. In particular, it has a property of many DDRSs: for a 
given object its $\alpha_{\rm inj}^{\rm inn}$ and $\alpha_{\rm inj}^{\rm 
out}$ are similar. Similarity of 
Q$_{\rm j}^{\rm inn}$ and Q$_{\rm j}^{\rm out}$ is also striking. These 
circumstances are in line with our conclusion that J0028$+$0035 is a 
{\em bona fide} DDRS which was so far not recognized as such only because
of the eastern component of the inner triple being a coincidence.

\item We found the age of the outer and the inner lobes of J0028$+$0035 -- 
245 and 3.6\,Myr, respectively -- and the length of quiescence between 
the two active phases -- 11\,Myr.

\item We argue that the inner lobes evolve into an almost uniform 
medium inside a cocoon surrounding them. Since it has been modified by the 
jet flow during the previous active phase of AGN, the power-law density 
distribution of the gaseous medium with a standard exponent $\beta\!=\!1.5$ 
cannot be adopted here. Otherwise, the obtained values of kinetic 
temperatures become unrealistic.

\end{enumerate}

\section*{Acknowledgements}

This work is (partly) based on data obtained with the ILT. LOFAR 
\citep{vanhaarlem2013} is the Low Frequency Array designed and constructed 
by ASTRON. It has observing, data processing, and data storage facilities in 
several countries, that are owned by various parties (each with their own 
funding sources), and that are collectively operated by the ILT foundation 
under a joint scientific policy. The ILT resources have benefited from the 
following recent major funding sources: CNRS-INSU, Observatoire de Paris and 
Université d'Orléans, France; BMBF, MIWF-NRW, MPG, Germany; Science 
Foundation Ireland (SFI), Department of Business, Enterprise and Innovation 
(DBEI), Ireland; NWO, The Netherlands; The Science and Technology Facilities 
Council, UK; Ministry of Science and Higher Education (MSHE), Poland. We 
thank the MSHE for granting funds for the Polish contribution to the ILT 
(MSHE decision no. DIR/WK/2016/2017/05-1) and for maintenance of the LOFAR 
PL-610~Borowiec, LOFAR PL-611~Łazy, and LOFAR PL-612~Bałdy stations. These 
data were (partly) processed by the LOFAR Two-Metre Sky Survey (LoTSS) team. 
This team made use of the LOFAR direction-independent calibration pipeline 
(https://github.com/lofar-astron/prefactor), which was deployed by the LOFAR 
e-infragroup on the Dutch National Grid infrastructure with support of the 
SURF Co-operative through grants e-infra 160022 and e-infra 160152 
\citep{Mechev2017}. The LoTSS direction-dependent calibration and imaging 
pipeline (http://github.com/mhardcastle/ddf-pipeline/) was run on compute 
clusters at Leiden Observatory and the University of Hertfordshire, which 
are supported by a European Research Council Advanced Grant 
[NEWCLUSTERS-321271] and the UK Science and Technology Funding Council 
[ST/P000096/1].\\
We thank the staff of GMRT that made these observations possible. GMRT is 
run by the National Centre for Radio Astrophysics of the Tata Institute of 
Fundamental Research.\\
The National Radio Astronomy Observatory running the VLA is a~facility of 
the National Science Foundation operated under co\-operative agreement by 
Associated Universities, Inc.\\
Funding for the Sloan Digital Sky Survey IV has been provided by the Alfred 
P. Sloan Foundation, the U.S. Department of Energy Office of Science, and 
the Participating Institutions. SDSS-IV acknowledges
support and resources from the Center for High-Performance Computing at
the University of Utah. The SDSS web site is www.sdss.org.
SDSS-IV is managed by the Astrophysical Research Consortium for the 
Participating Institutions of the SDSS Collaboration including the 
Brazilian Participation Group, the Carnegie Institution for Science, 
Carnegie Mellon University, the Chilean Participation Group, the French 
Participation Group, Harvard-Smithsonian Center for Astrophysics, 
Instituto de Astrof\'isica de Canarias, The Johns Hopkins University, Kavli 
Institute for the Physics and Mathematics of the Universe (IPMU) / 
University of Tokyo, the Korean Participation Group, Lawrence Berkeley 
National Laboratory, 
Leibniz Institut f\"ur Astrophysik Potsdam (AIP),  
Max-Planck-Institut f\"ur Astronomie (MPIA Heidelberg), 
Max-Planck-Institut f\"ur Astrophysik (MPA Garching), 
Max-Planck-Institut f\"ur Extraterrestrische Physik (MPE), 
National Astronomical Observatories of China, New Mexico State University, 
New York University, University of Notre Dame, 
Observat\'ario Nacional / MCTI, The Ohio State University, 
Pennsylvania State University, Shanghai Astronomical Observatory, 
United Kingdom Participation Group,
Universidad Nacional Aut\'onoma de M\'exico, University of Arizona, 
University of Colorado Boulder, University of Oxford, University of 
Portsmouth, 
University of Utah, University of Virginia, University of Washington, 
University of Wisconsin, 
Vanderbilt University, and Yale University.\\
The Pan-STARRS1 Surveys (PS1) and the PS1 public science archive have 
been made possible through contributions by the Institute for Astronomy, the
University of Hawaii, the Pan-STARRS Project Office, the Max-Planck Society
and its participating institutes, the Max Planck Institute for Astronomy,
Heidelberg and the Max Planck Institute for Extraterrestrial Physics,
Garching, The Johns Hopkins University, Durham University, the University
of Edinburgh, the Queen's University Belfast, the Harvard-Smithsonian
Center for Astrophysics, the Las Cumbres Observatory Global Telescope
Network Incorporated, the National Central University of Taiwan, the Space
Telescope Science Institute, the National Aeronautics and Space
Administration under Grant No. NNX08AR22G issued through the Planetary
Science Division of the NASA Science Mission Directorate, the National
Science Foundation Grant No. AST-1238877, the University of Maryland,
E{\"o}tv{\"o}s Lor\'and University (ELTE), the Los Alamos National Laboratory, and
the Gordon and Betty Moore Foundation.\\
MJ and JM were supported by Polish National Science Centre grant UMO-2018/29/B/ST9/01793.\\
We are grateful to the referee for a large number of useful suggestions.

\section*{Data availability}

The data underlying this paper will be shared on reasonable request to the 
corresponding author.


\label{lastpage}

\end{document}